\documentclass[apjl]{emulateapj}
\usepackage{graphicx}
\usepackage{pstricks}

\RequirePackage{natbib}

\begin{document}

\title{The Integrated Sunyaev-Zeldovich Effect as the Superior Method for Measuring the Mass of Clusters of Galaxies}

\author{Patrick M. Motl\altaffilmark{1}}
\author{Eric J. Hallman}
\author{Jack O. Burns}
\affil{Center for Astrophysics and Space Astronomy,
University of Colorado, Boulder, CO 80309}
\author{Michael L. Norman}
\affil{Center for Astrophysics and Space Sciences, University of
California, San Diego, 9500 Gilman Drive, La Jolla, CA 92093}
\altaffiltext{1}{Current Affiliation: Department of Physics and Astronomy, Louisiana State University, Baton Rouge, LA 70803}
\email{motl@rouge.phys.lsu.edu}
\slugcomment{Accepted to ApJ Letters}
\keywords{galaxies:clusters:general--cosmology:observations--hydrodynamics--methods:numerical--cosmology:cosmic microwave background}
\begin{abstract}
We investigate empirical scaling relations between the thermal
Sunyaev-Zeldovich effect (SZE) and cluster mass in simulated clusters
of galaxies.  The simulated clusters have been compiled from  four
different samples that differ only in their assumed baryonic physics. 
We show that the
strength of the thermal SZE integrated over a significant fraction of
the virialized region of the clusters is relatively insensitive to the
detailed heating and cooling processes in the cores of clusters by
demonstrating that the derived scaling
relations are nearly identical between the four cluster samples
considered.  For our synthetic
images, the central Comptonization parameter shows significant
boosting during transient merging events, but the integrated SZE
appears to be relatively insensitive to these events. Most
importantly, the integrated SZE closely
tracks the underlying cluster mass.
Observations  through the thermal SZE allow a strikingly
accurate mass estimation from relatively simple measurements that do
not require either parametric modeling or geometric deprojection and
thus avoid assumptions regarding the physics of the ICM or the
symmetry of the cluster. This result offers significant promise for
precision cosmology using clusters of galaxies.
\end{abstract}
\section{Introduction}
Observations of galaxy clusters are increasingly used to constrain cosmological parameters. Reliable
determinations of the cluster mass, abundance of clusters as a
function of redshift (e.g.\ \citet{haiman}), and gas fraction (e.g. \citet{vikh03}) lead to independent constraints
on $\Omega_M$, $\Omega_b$, and the dark energy equation of state parameter
$w$ (see e.g. \citet{pen}, \citet{sasaki}, \citet{allen}). In most cases, extraction of
cosmological parameters from clusters requires a high precision
measurement of cluster mass, often for a large sample of
clusters.  It is therefore critical to understand whether existing or
planned X-ray or SZE observations can yield sufficiently  precise
measurements to permit accurate determinations of
cosmological parameters. 

Historically, cluster gas mass is often deduced from  the X-ray
temperature using an empirical  relation
between mass and spectral temperature with relatively small scatter for high
mass clusters (e.g. \citet{sand}).  The
X-ray surface brightness of a cluster results from the
emission along the line of sight,  
\begin{equation}
S_X = \frac{1}{4\pi(1+z)^{4}} \int n_{e} n_H \Lambda(T) dl \propto \int n^2 T^{1/2} dl
\end{equation}
where the proportionality holds if $\Lambda$ includes only
bremsstrahlung emission; which is a reasonable approximation for massive
clusters of galaxies.  In reality, the emission function will include both
resonance lines and free-free emission.

In contrast, the Sunyaev-Zeldovich effect (SZE) arises from the 
inverse Compton scattering of cosmic microwave background photons
by electrons in the hot plasma of clusters of galaxies \citep{sze}. 
The strength of the thermal SZE is proportional to the Compton
parameter, $y$, which for non-relativistic electrons is essentially the integral of the
gas pressure through the cluster
\begin{equation}
   y = \int \frac{k_{B} T}{m_{e} c^{2}} \sigma_{T} n_{e} dl \propto \int n T dl.
\end{equation}
For very hot clusters, relativistic corrections must be included \citep{itoh}.
The central  value of the Compton $y$ parameter will be referred to
as $y_0$. We also consider integrals of the Compton parameter over
a finite projected radius; for example
$y_{500}$ is the integral of $y$ over a disk with of radius $r_{500}$
corresponding to an overdensity of 500 relative to the
critical density.

From the above discussion, these two cluster
observables  depend differently  on the state of the
cluster gas. The X-ray emission is more sensitive to the density, and
so one might expect X-ray observations to provide a better constraint
on the cluster mass.  However, this strong density dependence  means
that the emission is dominated by the cluster core, where the gas 
is subject to many complicating  effects, including both  cooling and
heating. 

The two methods also have very different selection biases. By its
nature, the SZE is redshift independent \citep{carl} and observations of
clusters through this window hold great promise to test cosmological models.  
Since the
thermal SZE measures the integral of the gas pressure and
assuming that the gas outside the core is in approximate hydrostatic
equilibrium within the dark matter potential, one expects
that SZE observations should provide robust measures of the
cluster mass - independent of the complicated interplay of
physical mechanisms that regulate the thermal state of the
cluster gas (e.g. \citet{blanch},\citet{benson}).

To investigate this expectation, we have constructed four 
catalogs of simulated clusters of
galaxies that differ only in the assumed input physics
for the baryonic component.
Each sample contains $\sim 100$ clusters
at the present epoch in the mass range from $
10^{14} \; M_{\odot}$ to $2 \times 10^{15} \; M_{\odot}$ and
$\sim 10$ clusters more massive than 
$10^{14} \; M_{\odot}$ at a redshift of 2, roughly corresponding to the expected
sensitivity of upcoming SZE telescopes.  

In this {\em Letter}, we use ideal synthetic observations of the thermal SZE and X-rays to
measure how cluster observables scale with cluster mass. Since no
instrumental effects are included, this study indicates the minimum
errors one can expect in these relations. Evaluating typical scaling
relations for clusters reveals that the integrated SZE is a
surprisingly accurate and simple method for determining the mass of
clusters. Thus, SZE observations  have a high potential for
precision measures of fundamental cosmological parameters.  
\section{Numerical Simulations}
Our simulations use the hybrid Eulerian adaptive mesh refinement/N-body code 
\textit{Enzo} (\citet{enzo}; http://cosmos.ucsd.edu/enzo)
to evolve both the dark matter and baryonic fluid in the clusters,
utilizing the piecewise parabolic method (PPM) for the
hydrodynamics. With up to seven levels of dynamic refinement
in high density regions, we attain spatial resolution up to
$\sim \; 16 \; h^{-1}$ kpc in the clusters. We assume a concordance
$\Lambda$CDM cosmological model with the following parameters:
$\Omega_{\mathrm{b}} = 0.026$, $\Omega_{\mathrm{m}} = 0.3$,
$\Omega_{\Lambda} = 0.7$, $\mathrm{h} = 0.7$, and $\sigma_{8} = 0.928$.
Refinement of high density regions is performed as described in \citet{motl04}. 
We have constructed a catalog of AMR refined clusters identified in the simulation
volume as described in \citet{loken}. The four catalogs
progressively include additional physical processes in the
calculation. Our baseline catalog was run with only adiabatic physics, a
second includes the effects of radiative cooling, a third adds a
model for the loss of low entropy gas to stars, and a final 
simulation adds a moderate
amount of supernova feedback to the star
formation prescription. 

The loss of energy to radiation
is calculated from a tabulated cooling curve derived from a Raymond-Smith
plasma emission model \citep{rs} assuming a constant metallicity of 0.3 relative to
solar. For the simulation runs including star formation, we
have incorporated the star formation prescription from \citet{cen}. 

Briefly, for every most refined cell, if the gas is contracting, cooling
rapidly, and contains more than a Jeans mass of material,
a new star particle is created with a mass
$m_{b} \; \eta \; \Delta t / t_{dyn}$ and this amount of mass is
removed from the fluid.  The star formation rate is thus coupled to
the local dynamical time, $t_{dyn}$, while $\eta$ parameterizes the
efficiency of star formation and $\Delta t$ is the simulation timestep
increment.  In the simulation including feedback from stars, the new
star particle begins to deposit energy in the fluid to simulate the
explosion of prompt, type II supernovae.
\begin{figure}
    \begin{center}
    \begin{tabular}{ccc}
     \large{\textbf{Temperature}} & \large{\textbf{X-ray}} & \large{\textbf{Thermal SZE}} \\
       \multicolumn{3}{c}{\large{\textit{Adiabatic}}} \\
       \includegraphics[scale=0.25]{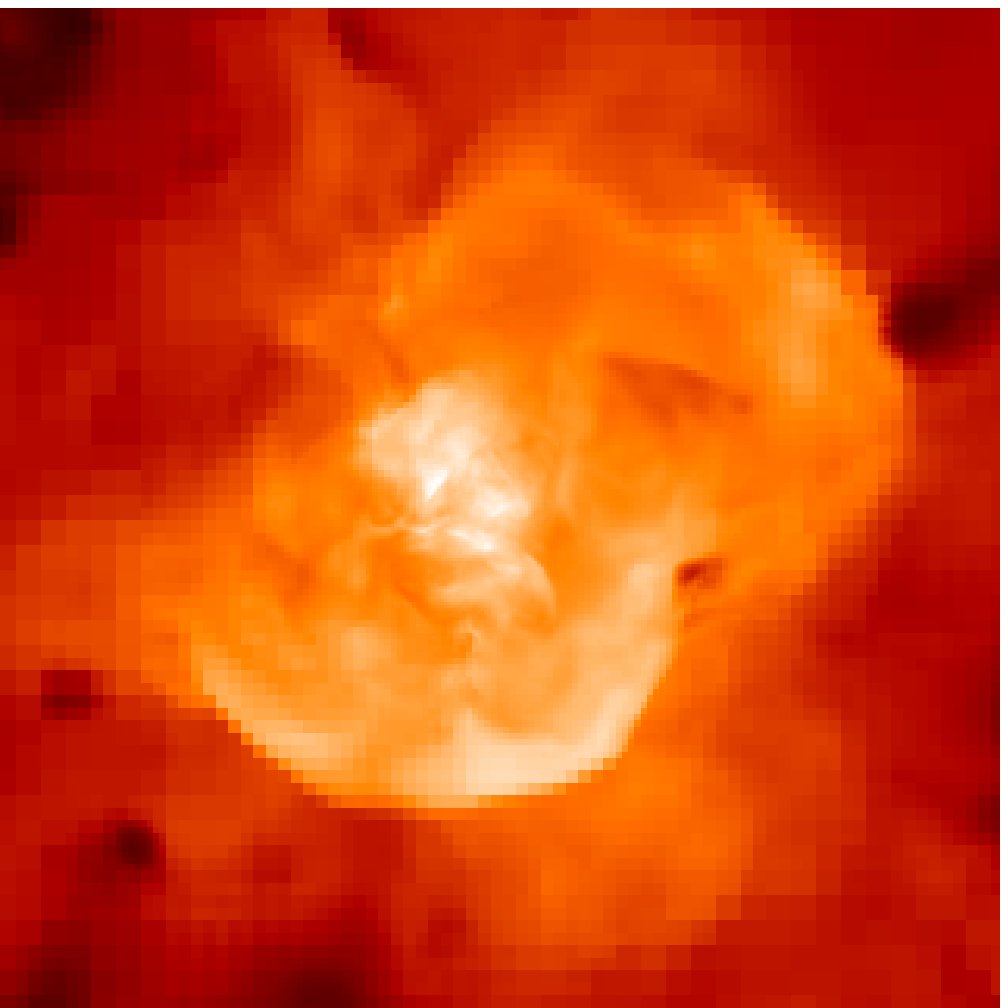} &
       \includegraphics[scale=0.25]{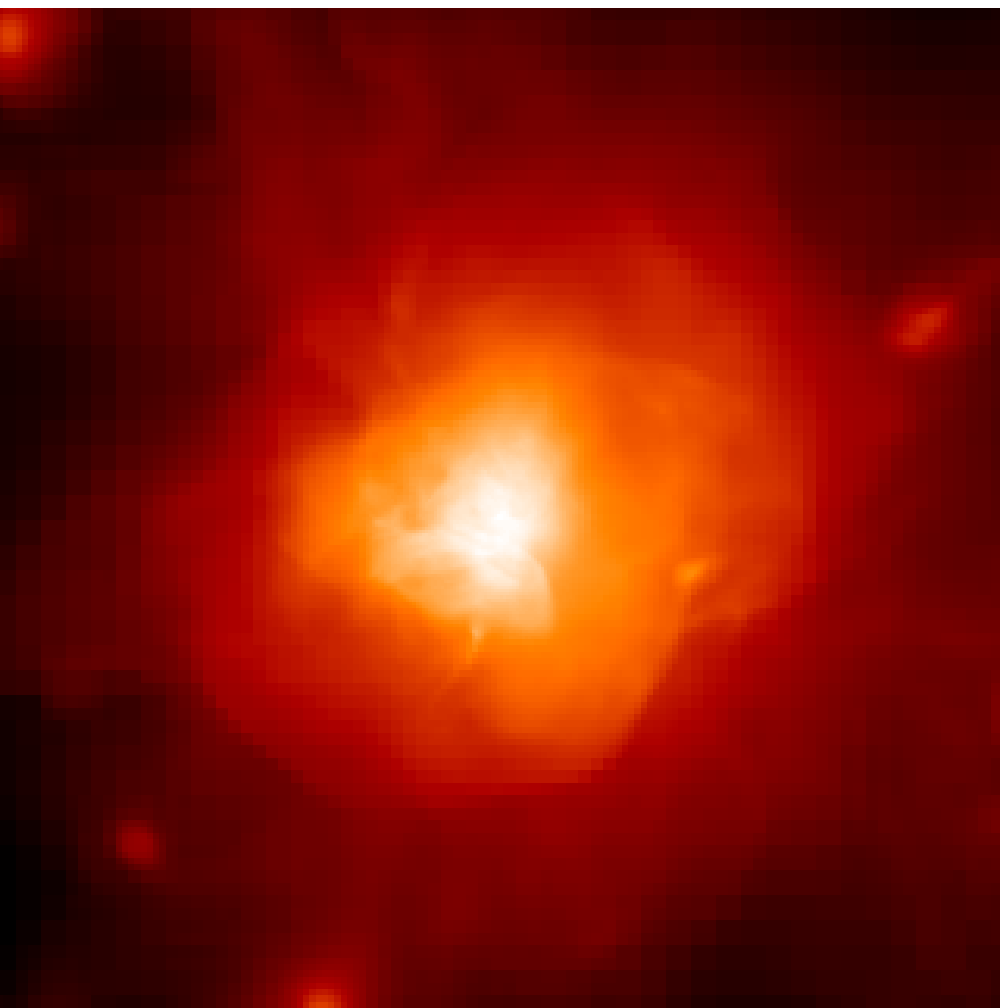} &
       \includegraphics[scale=0.25]{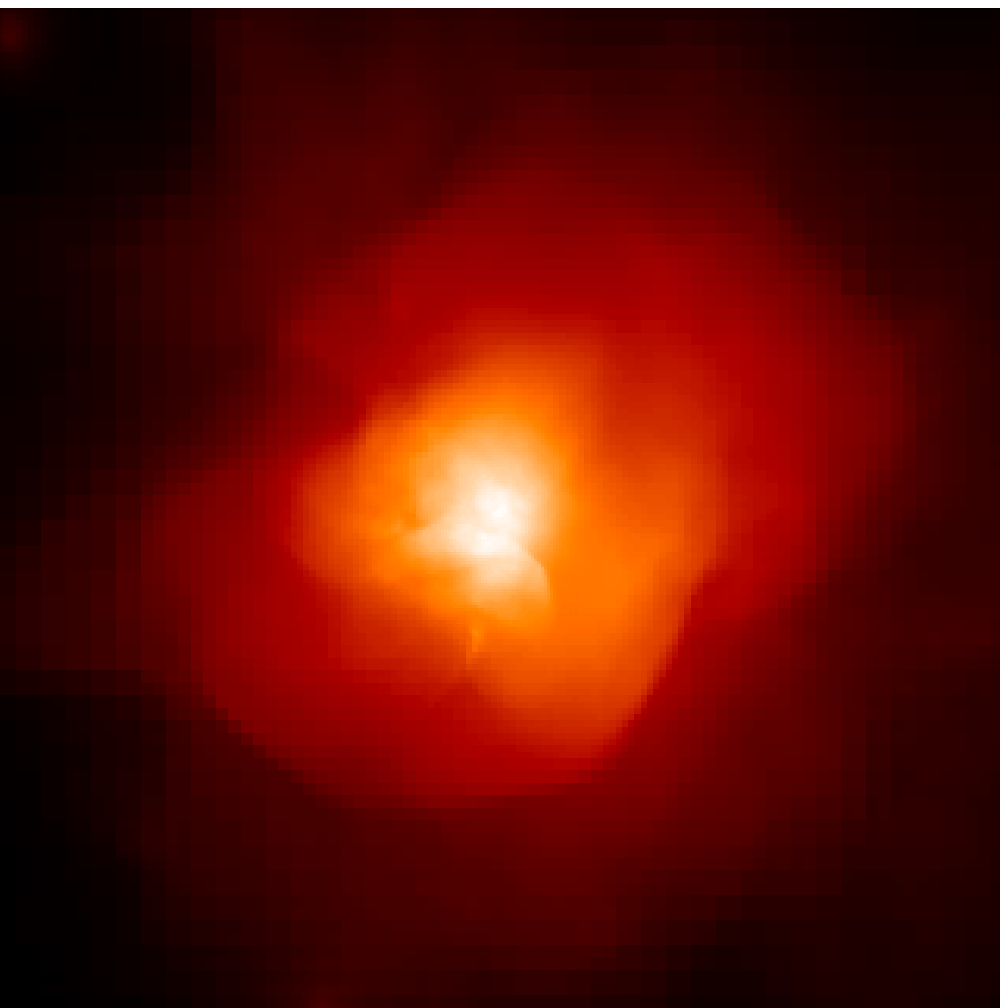} \\
       \multicolumn{3}{c}{\large{\textit{Radiative Cooling}}} \\
       \includegraphics[scale=0.25]{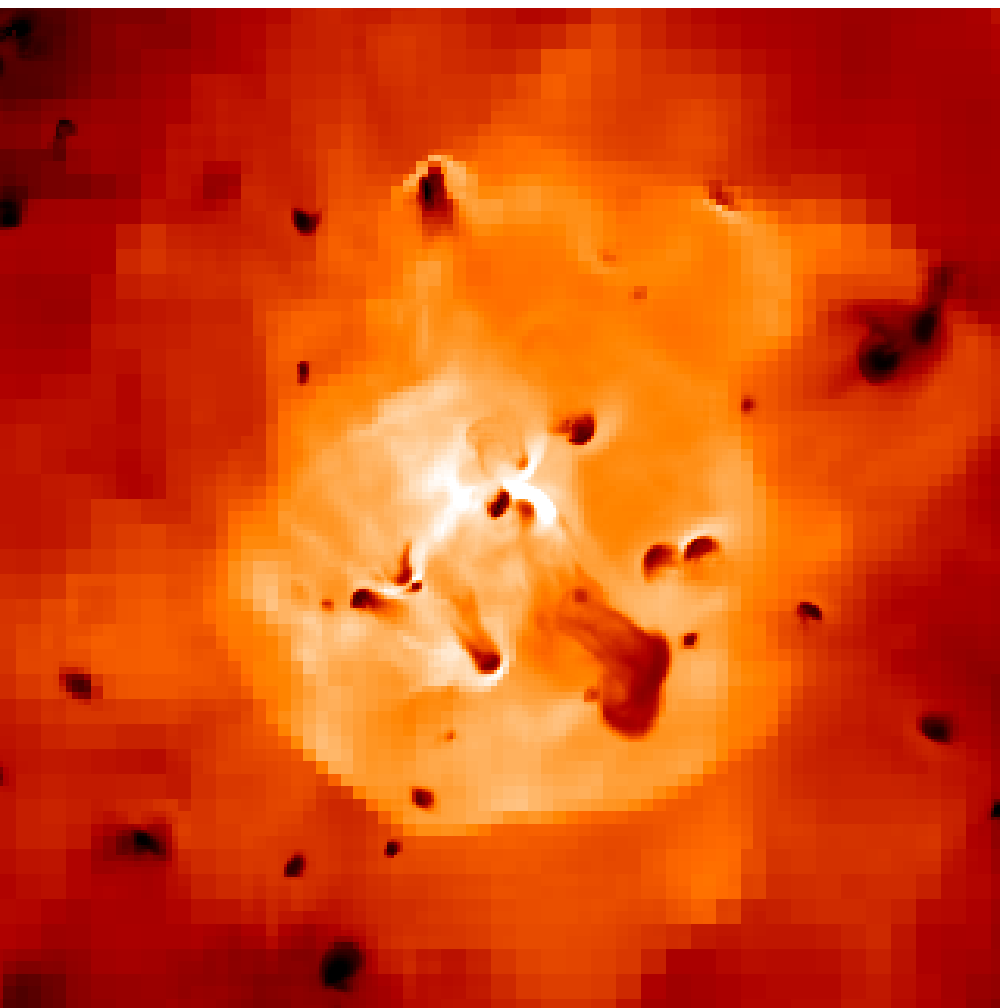} &
       \includegraphics[scale=0.25]{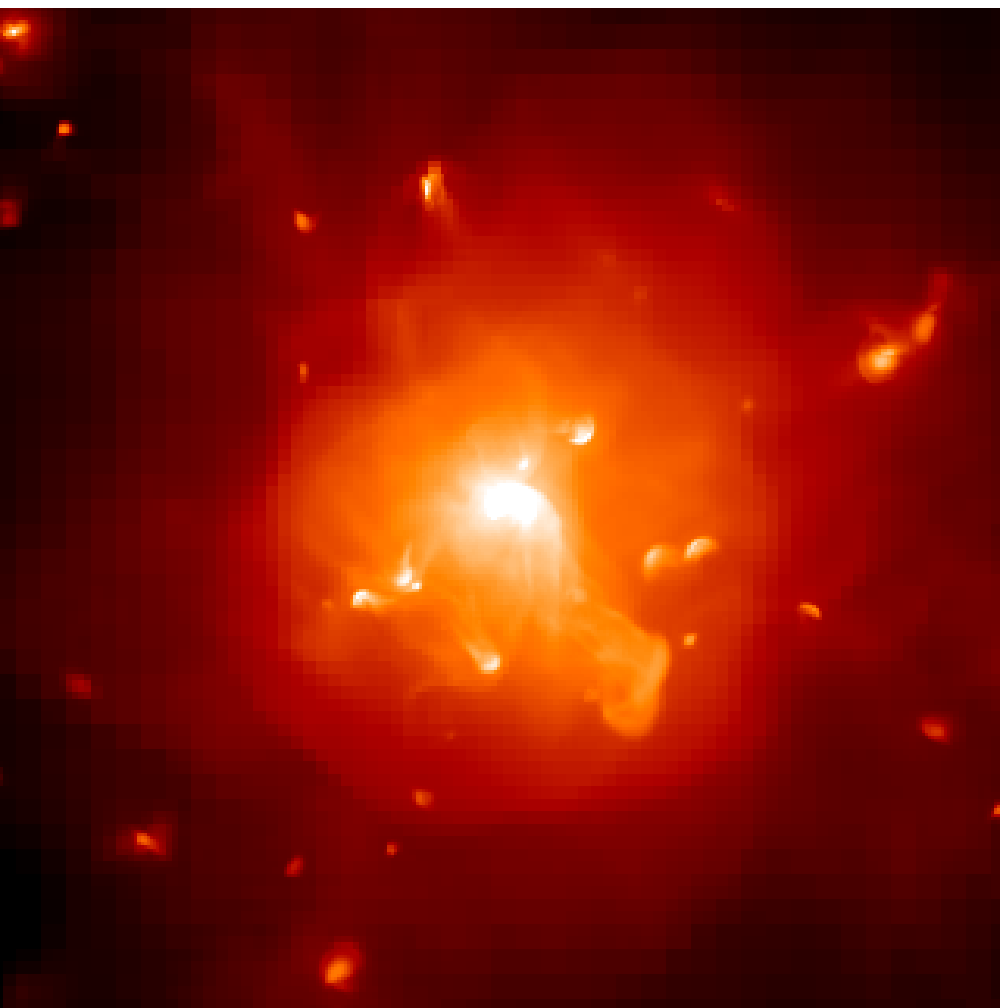} &
       \includegraphics[scale=0.25]{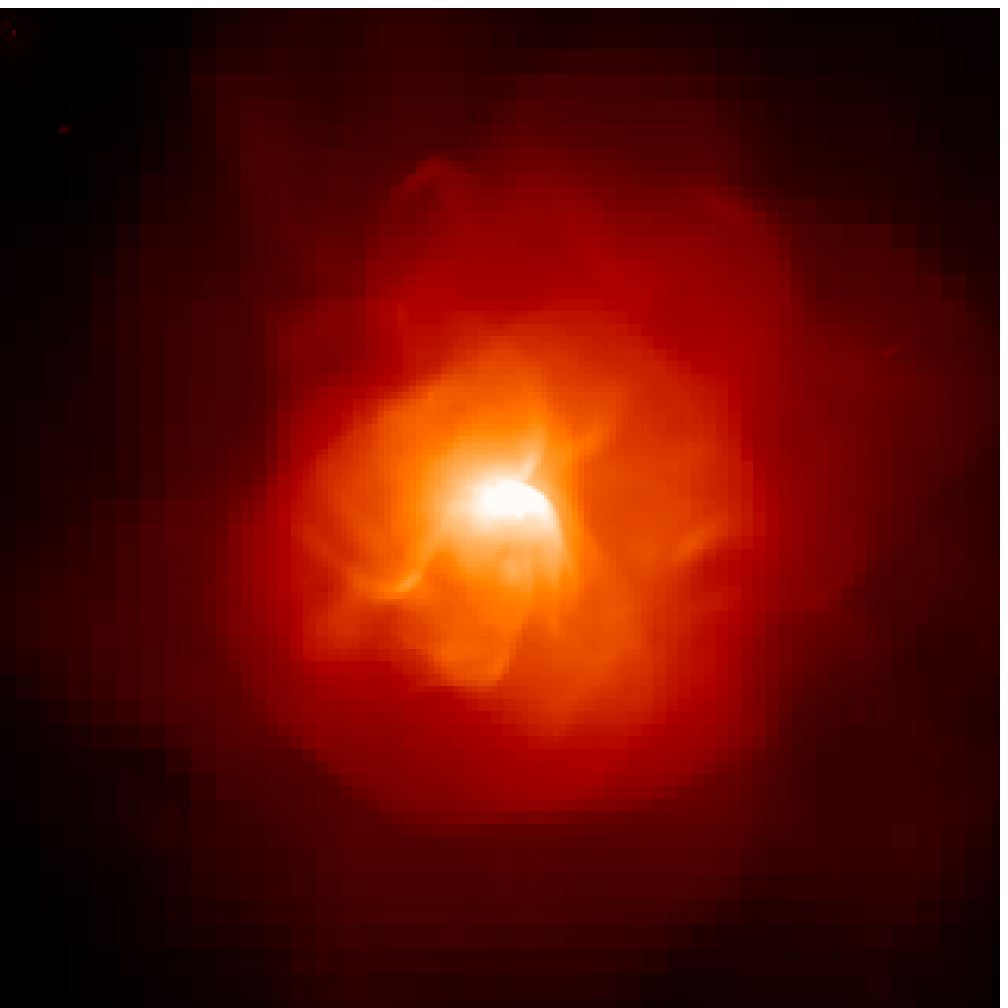} \\
       \multicolumn{3}{c}{\large{\textit{Star Formation}}} \\
       \includegraphics[scale=0.25]{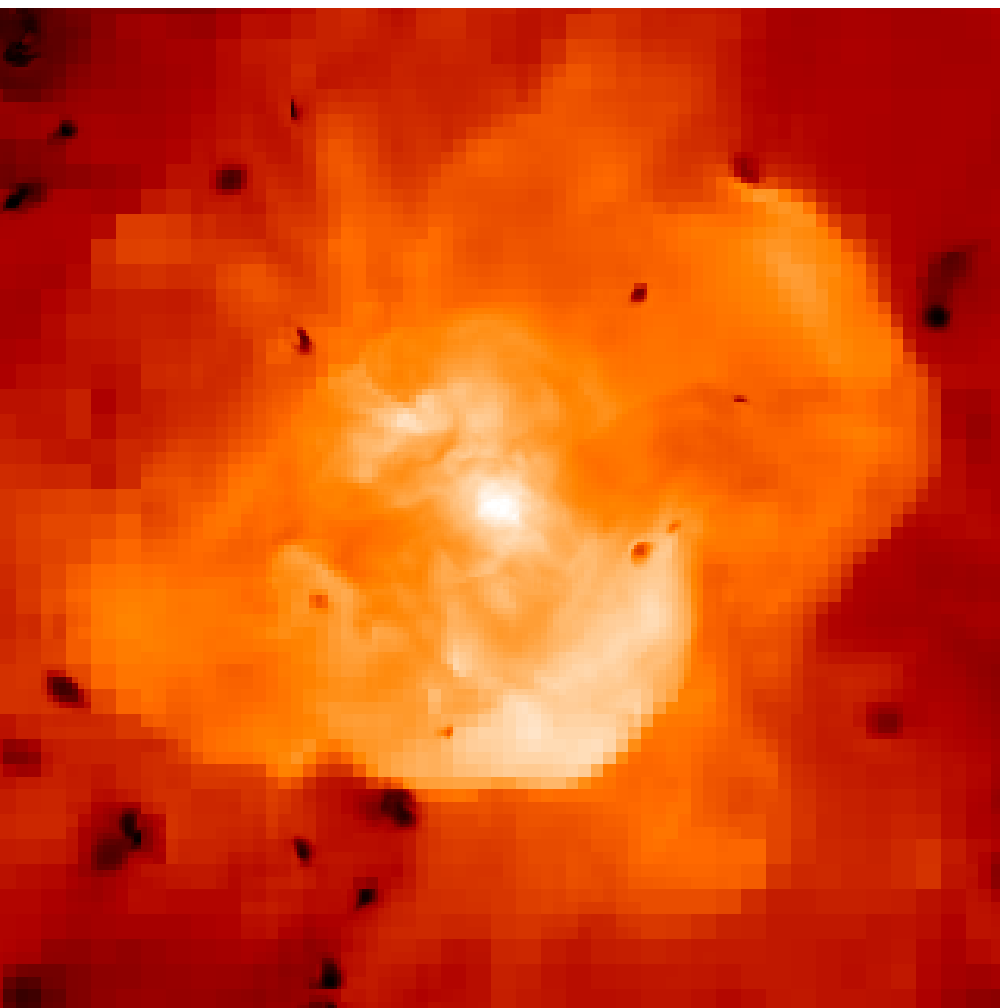} &
       \includegraphics[scale=0.25]{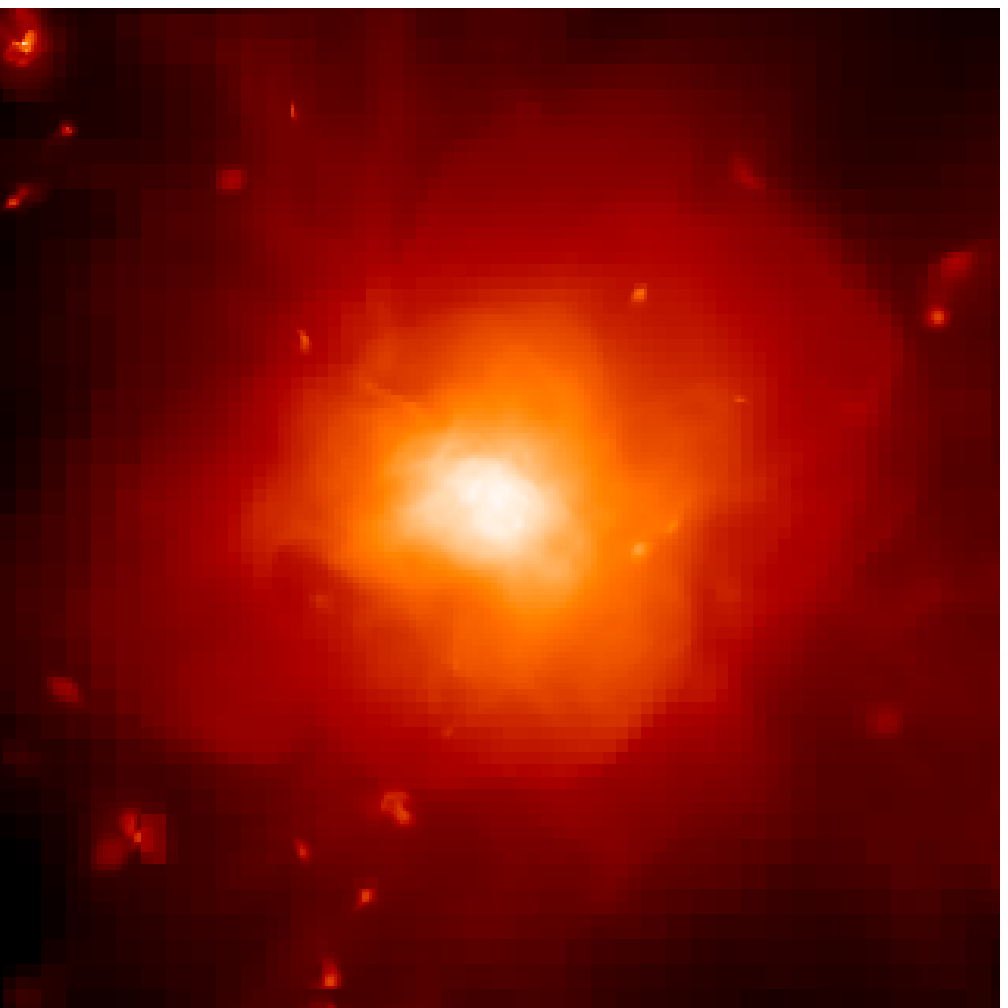} &
       \includegraphics[scale=0.25]{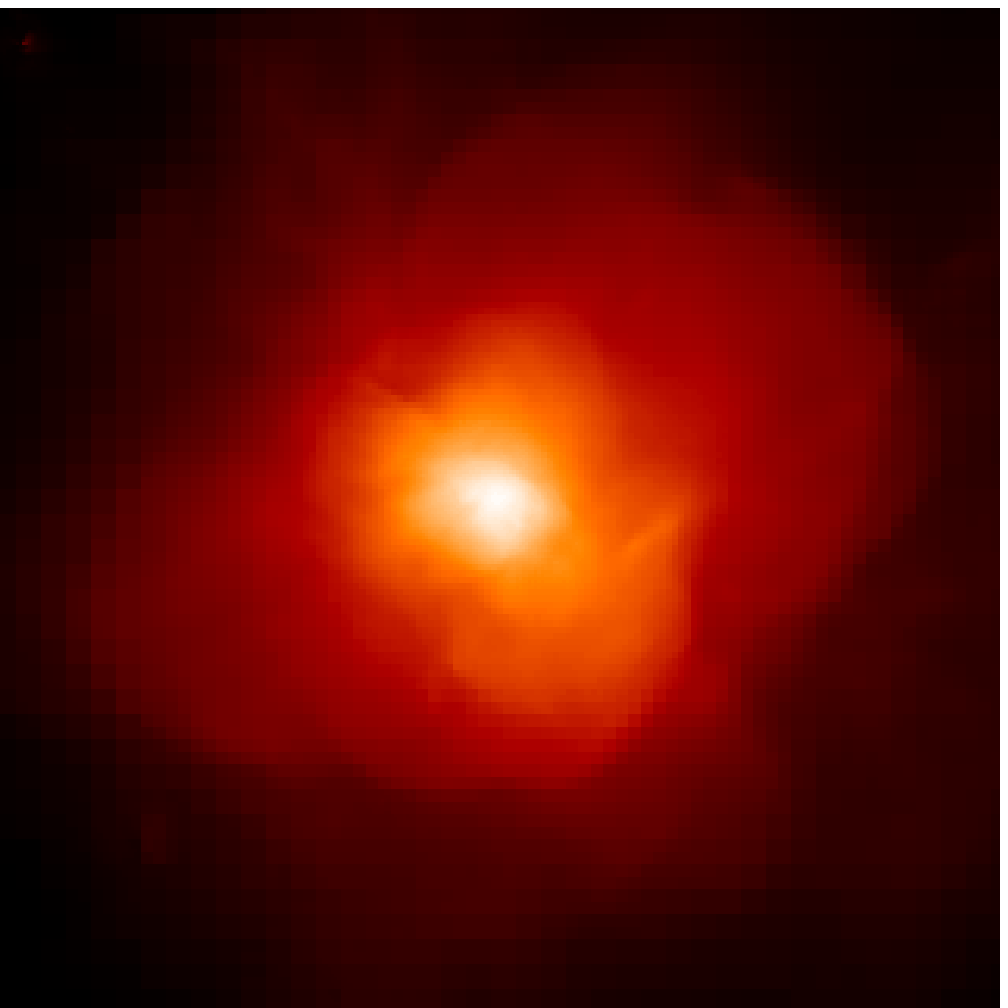} \\
       \multicolumn{3}{c}{\large{\textit{Star Formation Feedback}}} \\
       \includegraphics[scale=0.25]{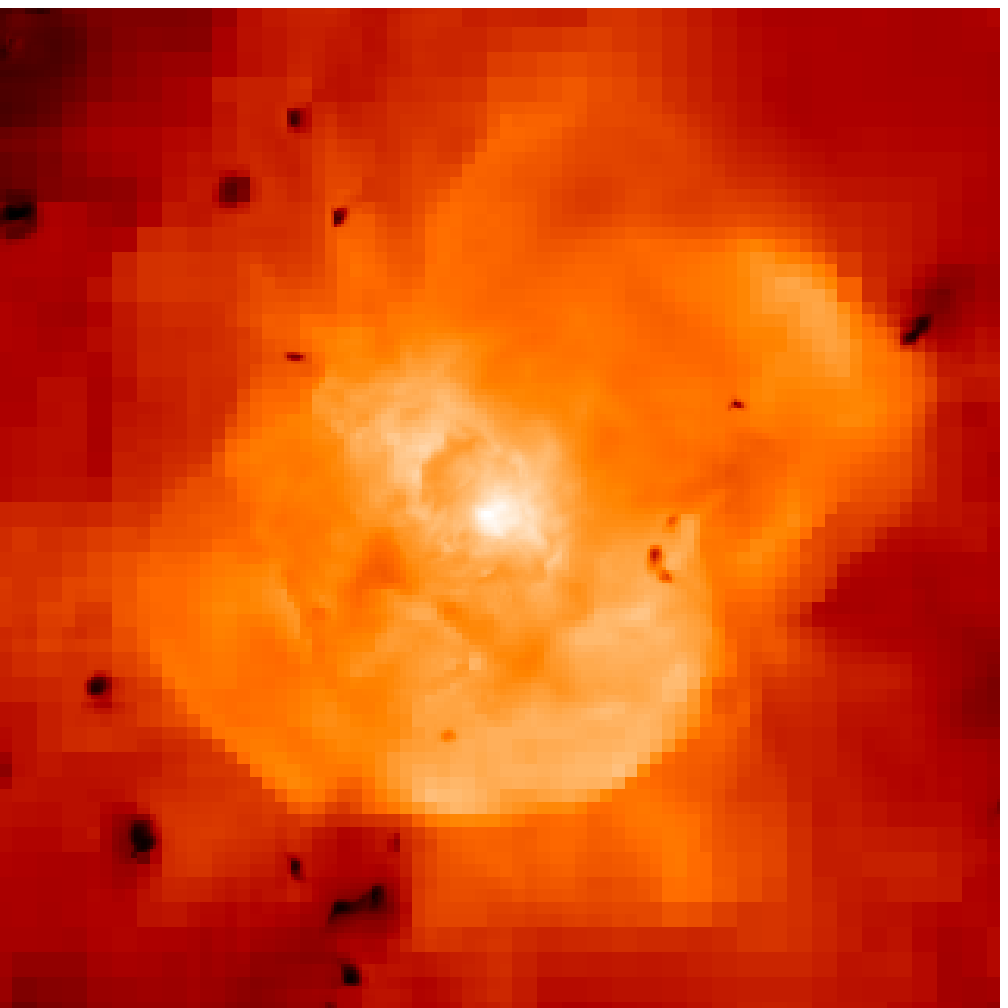} &
       \includegraphics[scale=0.25]{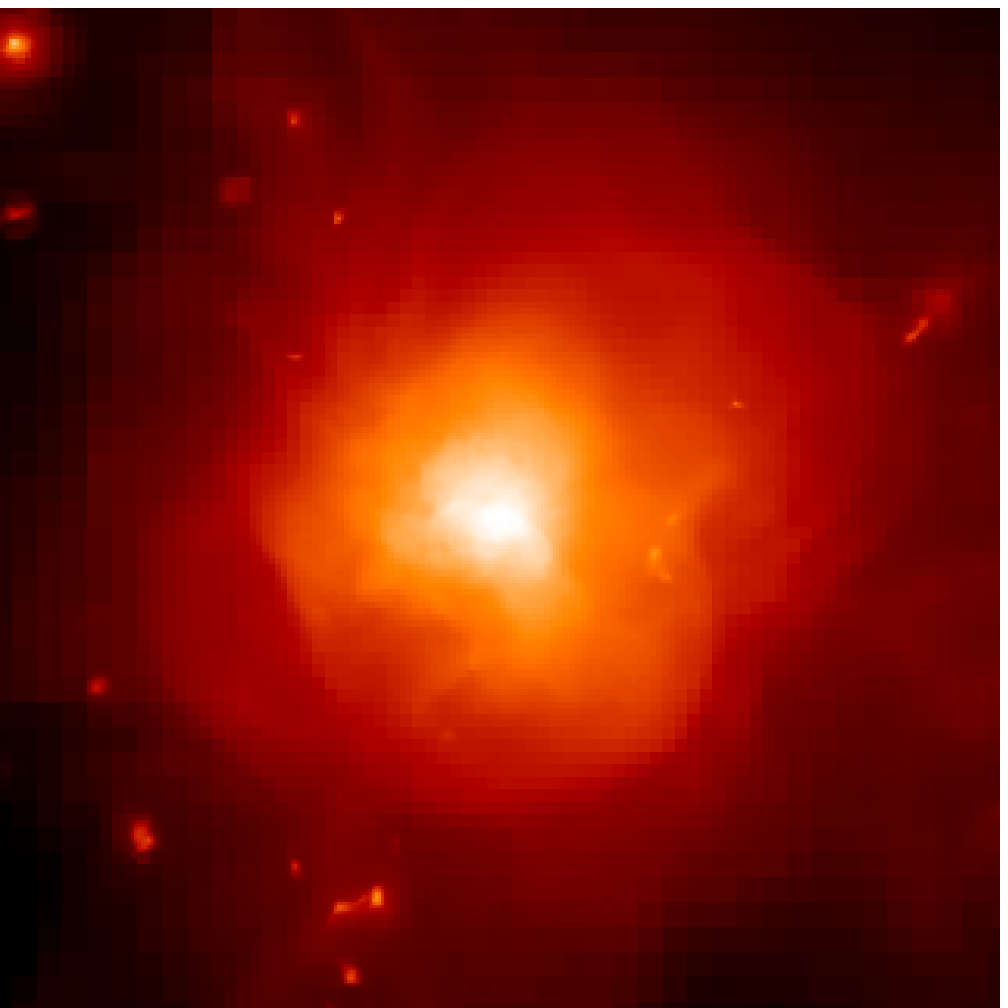} &
       \includegraphics[scale=0.25]{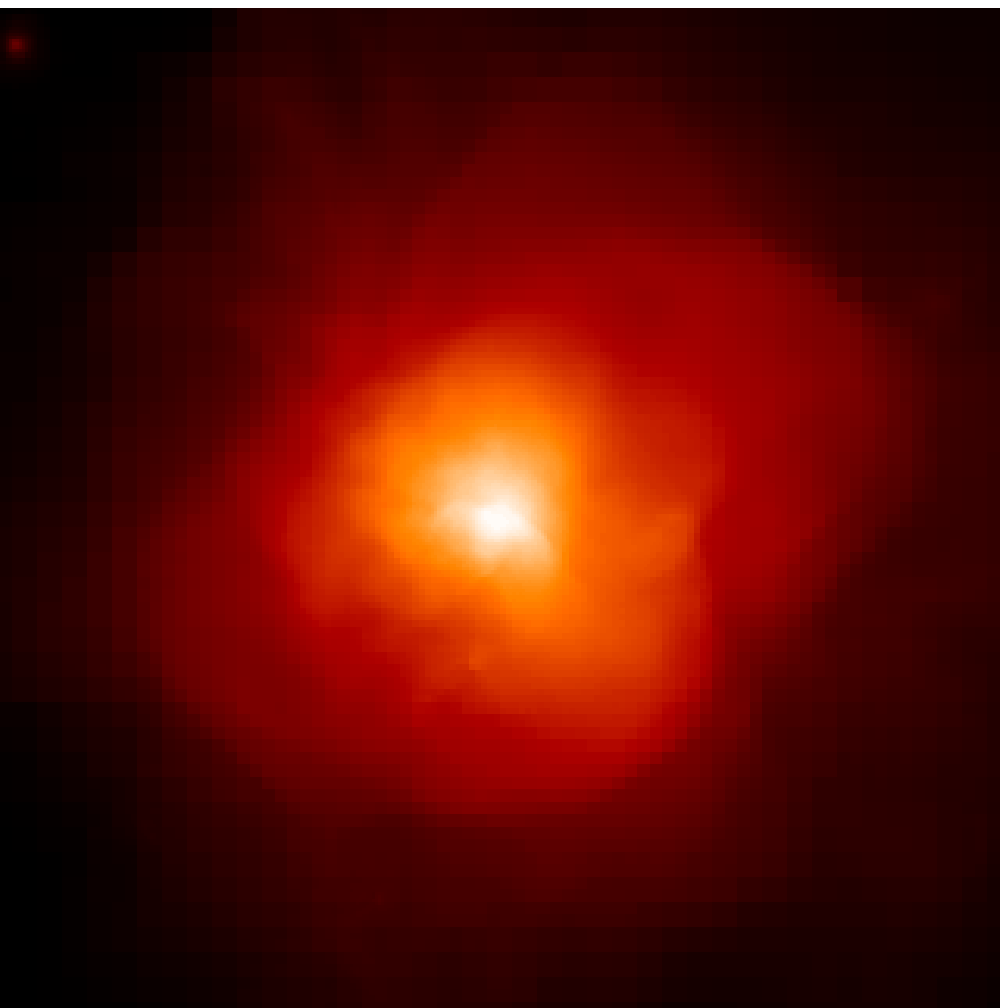} \\
       \includegraphics[scale=0.30]{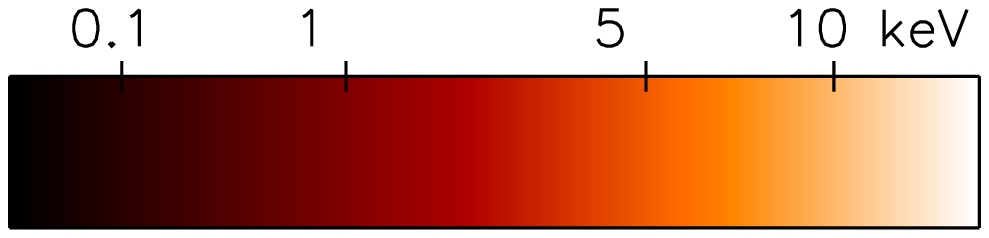} &
       \includegraphics[scale=0.30]{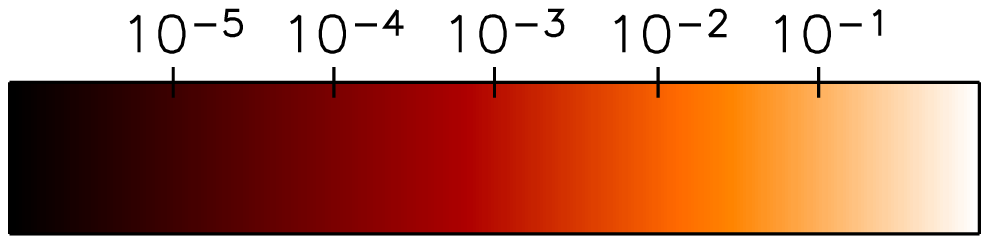} &
       \includegraphics[scale=0.30]{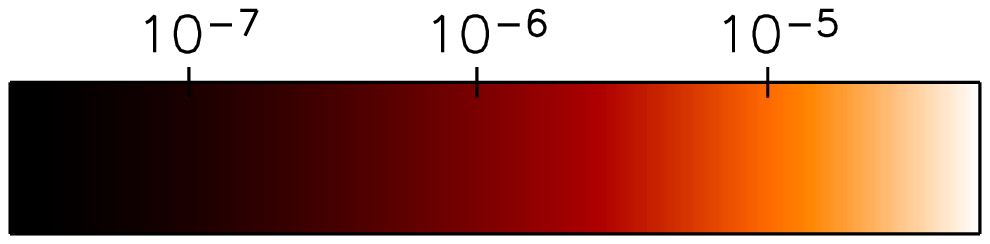} \\
    \end{tabular}
    \end{center}
   \vspace{-5mm}
   \caption{Images of the projected, emission-weighted temperature map (left column), normalized
    X-ray
    surface brightness (middle column) and the Compton parameter, $y$ (right column). 
    Each row displays the same cluster evolved in the
    indicated physical limit at the present epoch.  The field of view is $5 \; h^{-1} \; Mpc$ on a
    side.  The color table for each image is set by the range of values in the adiabatic realization.
    This particular cluster has a total mass, $M_{200} (\sim
    M_{virial}) \sim 2 \times 10^{15} \; M_{\odot}$. In
         reality, clusters should be intermediate between the limiting
         cases of radiative cooling only (which produces too many
         clusters with cool cores) and the star formation feedback
         model (which produces too few).  
     }
   \vspace{-6mm}
\end{figure}
\section{Results}
Figure 1 depicts typical simulation results from the four cluster
         samples that we have considered. The most dramatic difference
         between the samples is the  rich array of cool substructures in the cooling only simulation.  These cool cores and their associated shock features
    are very conspicuous in the temperature map and X-ray image.  While there is significant
         variation between clusters when viewed through their X-ray
         emission directly or through the derived temperature maps,
         the thermal SZE yields a quite similar appearance for the
         clusters - even in the extreme case of cooling only. 
\subsection{The Effect of Cluster Mergers}
The bias introduced by structure formation itself
         raises potential concerns, independent of the complexities of
         the detailed physics governing energy balance in the cluster
         gas.  Generally, we find that the central value of the Compton
         parameter ($y_0$) can be boosted by up to a factor of $\sim 20$ in
         head on collisions between equal mass subclusters.  These
         major mergers can drive the evolution of the central SZE signal over
         time periods of $1 - 2 \; t_{dyn}$ or $\sim 2 \; Gyr$.
         Furthermore, even minor mergers (which are more common) with
         a mass ratio of $\sim 0.1$ can produce significant boosts in
         $y_{0}$ by a factor of $\sim 2$.

	 On a cluster by cluster basis, merger boosts will bias
         measurements derived from the cluster properties to, for
         example, overestimate the mass of the cluster corresponding
         to a given SZE signal.  In a survey of clusters, merger
         boosts will scatter systems with equilibrium values
         below the detection threshold (which are more numerous) into
         the sample.

	 While mergers significantly impact $y_{0}$, 
	 they do not  impact the
	 value of $y_{500}$ as drastically. 
	 Measuring the integrated SZE out to a
	 significant radius has the effect of de-emphasizing
	  the core properties of the cluster, and so
	  the integrated SZE is  less sensitive to the dynamical
	 state of the cluster, as well as to the details of the
	 cluster physics as we show below. 
\begin{figure*}
\begin{center}
\pspicture(0,0.0)(0.0,9.5)
\includegraphics[scale=0.95]{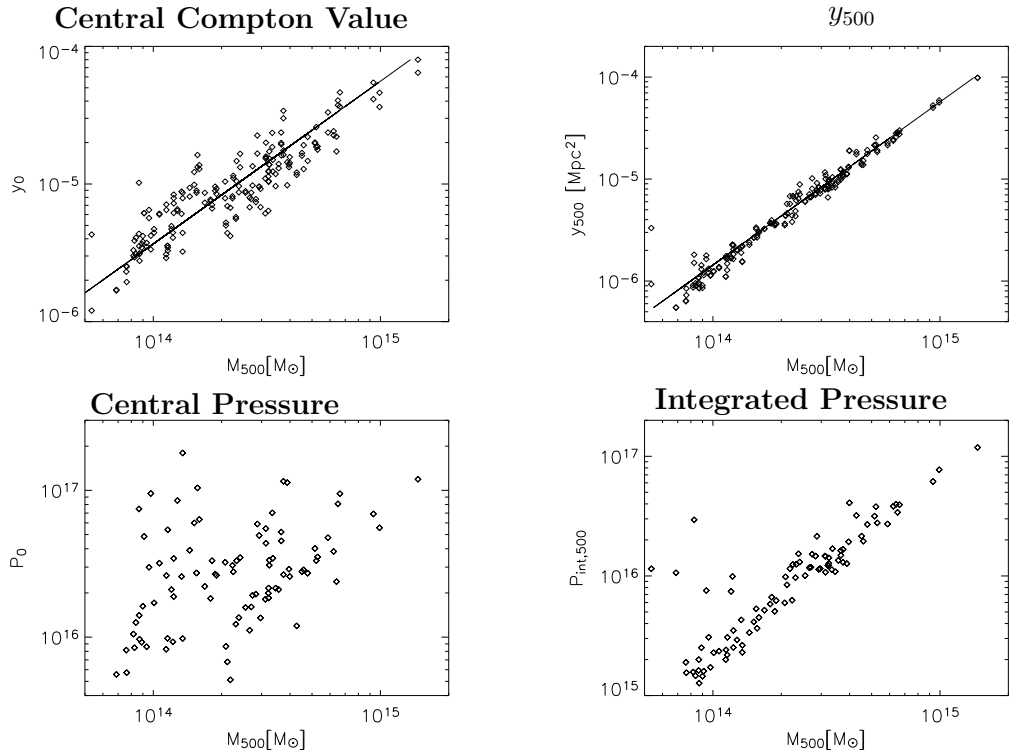}
\endpspicture
\end{center}
  \vspace{-6mm}
  \caption{Upper: The scaling relations between projected $y_0$ and $y_{500}$
   and the total cluster mass within the same radius at the present
   epoch for the star formation with feedback cluster sample.  Two
   randomly chosen, orthogonal projections for each cluster are
   plotted as individual points and the catalog contains $\sim 100$
   clusters at this epoch in the mass range $10^{14} \;
   M_{\odot} \le M_{200} \le 2 \times 10^{15} \; M_{\odot}$.  The best
   fit relations are plotted as solid lines.  Lower: Central pressure
   and pressure integrated inside sphere of radius $r_{500}$ plotted
   against cluster total mass. }
  \vspace{-6mm}
\end{figure*}
\subsection{Thermal SZE Scaling Relations}
	 The comparison of the $y_{500} - M_{500}$ relation with the
	 $y_0 - M$ relation is shown in the upper panels
	 of Figure 2 for the star formation with feedback sample.
	 The $y_0$ scaling shows a much larger scatter than
	 $y_{500}$, and so mass estimation using the central value
	 of the Compton parameter is correspondingly poorer.  To
	 illustrate the reason for this discrepancy, we show in the
	 bottom panels of Figure 2 both   the central
	 pressure in each cluster, and the
	 integrated pressure inside a sphere of radius $r_{500}$
	 as functions of
	 mass. $y_0$ is essentially a measure of the pressure integrated
	 along the line of sight
	 through the cluster center.
	 However, the integrated SZE,
	  $y_{500}$, measures the \textit{projected} integrated pressure
	 inside $r_{500}$. It is clear from these plots that while the
	 central pressure is weakly correlated with cluster mass, the
	 value of the pressure integrated out to a large radius is a very
	 good predictor of the mass. This difference is a result of
	 a variety of effects, including both mergers and non-adiabatic
	 physics,  which dominate
	 the cluster cores. Observing
	 the integrated SZE signal more effectively samples the cluster
	 potential depth by measuring the integrated pressure out to a
	 large radius, leading to a tight correlation of the strength of the
	 signal with cluster mass.  
\begin{table}
	   \begin{center}
	     \caption[ScalExp]{Scaling Exponent for $y_{500}-M$
	     Relation, z=0}
	 \begin{tabular}{ccc}
	   \tableline
	   \tableline
	   Simulation &  $\alpha$ & $\sigma_{\alpha}$ \\ 
	   \tableline
	   Adiabatic & 1.59 & 0.021\\
	   Radiative Cooling& 1.71 & 0.031\\
	   Star Formation& 1.60 & 0.027\\
	   Star Formation with Feedback& 1.61 & 0.024\\
	   \tableline
	   \end{tabular}
	 \end{center}
	   \vspace{-6mm}
	   \end{table}
	 Looking back in time through our catalogs of simulated
         clusters, we find that all four samples produce similar
         predictions for the scaling between $y_{500}$ and $M_{500}$. These
         predictions are roughly consistent with the expectations of
         self-similar scaling for clusters, up to a redshift
         of 1.5 for which we have a sufficient number of clusters
         in our simulations. Table 1 shows the scaling exponent
         and the variance for each of the four samples
         at $z =0$, where the scaling relation is written
         as
\begin{equation}
y_{500} = A \left[\frac{M_{500}}{10^{14} M_{\odot}}
  \right]^{\alpha}.
\end{equation}
 The values of the exponent $\alpha$ shown in the table are consistent with
         previous work by \citet{dasilva}. While the detailed
         simulation of clusters incorporating all relevant physics
         remains an unfulfilled challenge, it is reassuring that all
         models considered by us to date predict similar properties
         for the integrated SZE. This results from the weak dependence
         of the integrated SZE on the state of the cluster core.  Therefore, the
         interpretation and 
         modeling of SZE observations of clusters should be relatively
         immune to theoretical uncertainties and SZE surveys will
         provide excellent opportunities for constraining cosmological
         parameters. 
\begin{table}
	   \begin{center}
	     \caption[MassEst]{Accuracy of Mass Estimation, z=0}
	 \begin{tabular}{cccccc}
	   \tableline
	   \tableline
	   Method & Median $M_{est}/M_{true}$ & +1$\sigma$ &
	   -1$\sigma$ & +80\% & -80\% \\ 
	   \tableline
	   $y_{500}-M$ & 0.97 & 1.00 & 0.93 & 1.13 & 0.86\\
	   $T_X - M$& 1.04 & 1.11 & 0.92 & 1.33 & 0.74\\
	   $L_{X,500}-M$& 0.87 & 1.03 & 0.75 & 1.46 & 0.62\\
	   $y_0 - M$ & 0.96 & 1.14 & 0.82 & 1.53 & 0.67\\
	   \tableline
	   \end{tabular}
	 \end{center}
	   \vspace{-8mm}
	   \end{table}      
\subsection{Cluster Mass Estimates}
The relationship between X-ray spectral temperature and cluster mass
	 ($M-T_X$)
	 is well documented and a similar relation
	 is noted in
	 cosmological simulations (e.g., \citet{bryan}). 
	This relationship is often used to
	 infer cluster masses. For comparison to the $y_{500}-M$
	 relation studied here, we have measured several other scaling
	 relations, including $M-T_X$, for the simulated clusters.$T_X$ is in this case calculated as the
	 X-ray emission weighted average temperature inside a
	 projected radius of $r_{500}$.  
	 We then use the best-fit scaling relations derived from the
	 star formation with feedback simulation to
         determine the total cluster mass. Table 2 shows the ratio of
         estimated mass to true mass for four different scaling
         relations at z=0. 
	 
	 We find that the integrated
         thermal SZE signal within a characteristic radius of
         $r_{500}$ is the best estimator of the true cluster mass
	 among these scaling relations.  For example, 80\% of
         clusters  have estimated
         masses that lie within $+ 15\%$ to $-10\%$ of the true
         cluster mass and this accuracy is nearly constant with
         redshift back to at least $z = 1.5$.  The $y_{0}$ value is a much
         poorer estimate of the true cluster mass, as is expected from the
         relative scatter of $y_0 - M$ shown
         in Figure 2.  Core effects
	 also have a
	 strong impact on the X-ray luminosity, because of its
	 enhanced density dependence. Therefore the scaling of X-ray
	 luminosity with cluster mass is the poorest of these scaling
	 relations. 

	 The $M - T_{X}$
         relation for clusters in the simulation generates relative
         errors of $\pm$30\% in mass for 80\% of
         clusters. \citet{rasia} show that, from simulations, a
         $1\sigma$ scatter of 30\% in the $M-T_X$ relation is reduced
         to 16\% when using their spectroscopic-like temperature in
         place of the emission weighted temperature.  A similar drop in the
         scatter of the mass estimates quoted here for 80\% of the
         clusters would make the $M-T_X$ relation roughly equivalent to
         $y_{500}-M$ in predictive power. However, this analysis
         includes no instrumental effects; a detailed study of the
         additional error introduced  using real
         observations remains to be done in order to gauge the
         effectiveness of these methods. Certainly, measuring the
         value of $y_{500}$ requires no assumptions about the state of
         the cluster gas, nor any model fitting, whereas X-ray
         temperature is generally fit to a model assuming uniform
         temperature and metallicity within the gas. 
         SZE observations also can be done for
         a larger number of clusters, and at higher redshifts, than
         X-ray observations. In any case, of the 
         methods tested here, the integrated SZE scaling relation is
         superior at estimating cluster masses.
\section{Conclusions}
The integrated SZE in clusters of galaxies is not strongly
dependent on the details of cluster physics or variation in the cluster core in
contrast to the X-ray luminosity. 

While the value of $y_0$ experiences strong ``boosting'' due to
mergers, the value of $y_{500}$ appears to be much less sensitive to
transient  events. Measuring the SZE out to larger radii has
the effect of smoothing out the impact of mergers. 

The $y_{500} - M$ relation is the simplest, least model dependent, and
most accurate measure of cluster mass. An unprecedented number of
clusters will be observed through the SZE in the near future. Clusters
observed through the SZE therefore show great promise as precision
cosmological probes.   
\acknowledgments
The simulations presented in this work were conducted at the
National Center for Supercomputing Applications at the University of
Illinois, Urbana-Champaign through computer allocation grant
AST010014N.  We wish to acknowledge the support of the Chandra X-ray
science center and NASA through grant TM3-4008A.  We also acknowledge the
support of the NSF through grant AST-0407368.

\bibliographystyle{apj}
\bibliography{ms.bbl}

\end{document}